\documentstyle[preprint,aps,tighten,epsfig]{revtex}

\begin{document}

\draft

\title{THREE-PION INTERFEROMETRY
OF RELATIVISTIC NUCLEAR COLLISIONS}

\author{Hiroki Nakamura$^{(1)}$ and Ryoichi Seki$^{(2,3)}$}

\address{
${}^{(1)}$ Department of Physics,
Waseda University, Tokyo 169-8555, Japan \\
${}^{(2)}$ Department of Physics, California State University,
Northridge, CA 91330 \\
${}^{(3)}$ W.~K.~Kellogg Radiation Laboratory 106-38, 
California Institute of Technology, \\
Pasadena, CA 91125} 

\date{\today}

\maketitle

\begin{abstract}
Three-pion interferometry is investigated for 
new information on the space-time structure of
the pion source created in ultra-relativistic heavy-ion collisions.
The two- and three-pion correlations are numerically computed 
for incoherent source functions based on the Bjorken hydrodynamical model, 
over a wide range of the kinematic variables.
New information provided by three-pion interferometry, 
different from that provided by two-pion interferometry, 
should appear in the phases of the Fourier transform of the source function. 
Variables are identified that would be sensitive to the phases and 
suitable for observation.  For a positive, chaotic source function, however, 
a variation of the three-pion phase is found to be difficult to extract  
from experiments.  Effects of asymmetry of the source function are also 
examined.
\end{abstract}

\bigbreak

\section{Introduction}

Two-pion interferometry has been regarded as an important means of obtaining 
the space-time structure of dynamics involved in relativistic heavy-ion 
collisions.  Over the years, extensive studies of two-pion interferometry 
have been carried out theoretically and experimentally to investigate 
how much information the correlations of two emitted pions can provide.  
As experiments become refined, measurements of three-pion correlations 
should become feasible, hopefully providing new information.
The first result of such measurements has been recently reported 
from the CERN NA44 experiment\cite{exp}. 

In the last few years, some theoretical investigations have been made 
regarding three-pion interferometry\cite{HeiZha,CramerKdija,HeiVis,Humanic}.
Though the treatments of three-pion interferometry and the issues 
involved are similar to those regarding two-pion interferometry, 
three-pion interferometry is technically far more complicated than 
the two-pion case due to the involvement of an additional momentum and 
also to new aspects of the particle correlations.  Consequently, 
theoretical work so far has been on limited aspects of the interferometry, 
focusing on a few kinematic variables, such as the sum of three relative 
invariant momenta, over small ranges of their values. 

In the coming years, especially when RHIC becomes in operation, we expect 
that experiments on three-pion interferometry will become more detailed 
and will be made over larger ranges of various kinematic variables.
We report here an investigation of three-pion interferometry over 
a wide range of kinematic variables.  Our major objective is 
to clarify how much new information we can extract 
from three-pion interferometry, regarding the space-time structure of 
the pion source created in ultra-relativistic heavy-ion collisions. 
This work is similar to that of Heinz and Zhang\cite{HeiZha} in its 
objective, but it differs in scope.
We numerically calculate the two- and three-pion correlations 
for the same source functions and compare the two types of correlations. 
The calculations are carried out with various model source functions 
that are based on the Bjorken hydrodynamical model.  Throughout this work, we 
assume pion emission from the source to be completely chaotic.  We also 
neglect possible final-state interactions (including the Coulomb 
interactions) between the emitted pions and the source.  

The new aspect of the correlations that three-pion interferometry 
can provide is the phase of the source function's Fourier transform.  
The information content of the phase differs 
from that of the magnitude of the Fourier transform.  Since two-pion  
interferometry can provide only the magnitude, we hope that three-pion 
interferometry can provide new information.  To this end, we carefully 
identify the variables that are sensitive to the phase and suitable for 
observation.  Since the phase is expected to be greatly affected 
by asymmetry of the source function, we also examine that issue.

In Sec. II, we summarize the formulation of both the two- and the three-pion 
interferometries, mainly to define the correlation functions used.  
In Sec. III, we describe the various source functions that are used in the 
calculation, based on the Bjorken hydrodynamical model.
The choice of optimal variables in the three-pion correlations is  
given in Sec. IV, and various numerical results are described in Sec. V.
Discussion and conclusions are presented in Sec. VI.

\section{Formalism}

The two- and three-particle correlation functions that we discuss are 
reasonably well-known, but for clarity we sketch the formalism of, in our case, 
bosons, and we define the correlation functions and various variables that 
appear in this work.

The field operator $\varphi(x)$ of emitted bosons, such as pions and kaons, 
obeys the Klein-Gordon equation,
\begin{equation}
(\partial^2+m^2)\varphi(x) = J(x),
\end{equation}
where $m$ is the mass of the particle, and $J(x)$ the source
current\cite{notation}.
The incoming and outgoing states of $\varphi(x)$ are specified in terms of
the creation and annihilation operators $a^\dagger$ and $a$ for these states.
They are related to $J(x)$ as
\begin{equation}
a_{\rm{out}}({\bf{p}}) = a_{\rm{in}}({\bf{p}}) 
  + i \int d^4x\frac{1}{ \sqrt{(2\pi)^3\cdot 2\omega_p}} J(x) e^{-i p\cdot x}, 
\end{equation}
where $p_0$ is on-shell and $\omega_p = \sqrt{{\bf{p}}^2+m^2}$.

We define an important quantity in this work, a source function, $S(x,K)$, 
in terms of $J(x)$: 
\begin{equation}
S(x,K) = 
    \int \langle J^\dagger(x+\frac{y}{2})J(x-\frac{y}{2}) \rangle
    e^{i K\cdot y} d^4y,
\end{equation}
where $\langle ... \rangle$ denotes the quantum ensemble average over various 
incoming states.  The Fourier transform of $S(x,K)$ is defined as\cite{HeiZha}
\begin{eqnarray}
\rho (q,K) &=& \int S(x,K) e^{i q \cdot x} d^4x \\
         &\equiv& f (q,K) \exp( i \phi (q,K)).
\end{eqnarray}
where $f (q,K)$ is the amplitude and $\phi (q,K)$ the phase. 
For a real $S(x,K)$, we have $f (q,K) = f (-q,K)$ and 
$\phi (q,K) = - \phi (-q,K)$, and thus $\phi (0,K) = 0$. 
Boson spectra and correlation functions will be expressed in terms of  
$\rho (q,K)$ or $S(x,K)$, and  
$\phi (q,K)$ is written in the small-momentum expansion as\cite{HeiZha}
\begin{equation}
\phi (q,K) = \langle x \rangle_K \cdot q  
-\frac{1}{6}\left\langle (q \cdot \tilde{x}_K )^3 \right\rangle_K  
+ O(q^5),
\label{exphi}
\end{equation}
where $\langle ... \rangle_K = \int S (x,K) ... d^4x$, and 
$\tilde{x}_K = x - \langle x \rangle_K$.
Since $\phi (q,K)$ is written in terms of odd moments of $x$, 
an asymmetric (about $x$) source causes a strong, nonlinear dependence 
of $\phi (q,K)$ on $q$.
In this work, we examine the $S(x,K)$ of simple forms as discussed in Sec. III.

The one-particle spectrum is given by 
\begin{eqnarray}
W_1({\bf{p}}_1) &=& \langle a^\dagger_{\rm{out}}({\bf{p}}_1) 
                    a_{\rm{out}}({\bf{p}}_1) \rangle  \nonumber \\
 &=& \frac{1}{(2\pi)^3 2\omega_{p_1}} \int S(x,p_1) d^4x \nonumber \\
 &=& \frac{1}{(2\pi)^3 2\omega_{p_1}} f (0,p_1).
\end{eqnarray}
Higher-order correlations of $J(x)$ are assumed to satisfy Gaussian 
reduction, such as 
\begin{eqnarray}
\lefteqn{
\langle J^\dagger(x_1) J^\dagger (x_2) J(y_1) J(y_2) \rangle
 =} \nonumber \\
&& \langle J^\dagger(x_1) J(y_1) \rangle
   \langle J^\dagger(x_2) J(y_2) \rangle
 + \langle J^\dagger(x_1) J(y_2) \rangle
   \langle J^\dagger(x_2) J(y_1) \rangle.
\end{eqnarray}
The two-particle spectrum is then written as
\begin{eqnarray}
W_2({\bf{p}}_1,{\bf{p}}_2) &=& \langle
   a^\dagger_{\rm{out}}({\bf{p}}_1) a_{\rm{out}}({\bf{p}}_1) 
   a^\dagger_{\rm{out}}({\bf{p}}_2) a_{\rm{out}}({\bf{p}}_2)
   \rangle  \nonumber \\
 &=& W_1({\bf{p}}_1)W_1({\bf{p}}_2) 
    +\frac{1}{(2\pi)^6\cdot 2\omega_{p_1} \cdot 2\omega_{p_2} }
        \left|\int S(x,K_{12}) e^{i q_{12} \cdot x} d^4x \right|^2
 \nonumber \\
 &=& W_1({\bf{p}}_1)W_1({\bf{p}}_2)
    +\frac{1}{(2\pi)^6\cdot 2\omega_{p_1}\cdot 2\omega_{p_2}}
       f^2(q_{12},K_{12}),
\end{eqnarray}
where for the pair of the $i$-th and $j$-th momenta, we define their average 
and relative momenta as 
\begin{eqnarray}
K_{ij} &=&\frac{{p}_i + {p}_j}{2}, \nonumber \\
q_{ij} &=& {p}_i - {p}_j,
\end{eqnarray}
respectively.  The two-particle correlation function is then expressed as
\begin{eqnarray}
\label{eq:c2}
C_2({\bf{p}}_1,{\bf{p}}_2) 
&=& \frac{W_2({\bf{p}}_1,{\bf{p}}_2)}
         {W_1({\bf{p}}_1)W_1({\bf{p}}_2)} \nonumber \\
  &=& 1 + \frac{f_{12}^2}{f_{11}f_{22}}.
\end{eqnarray}
Hereafter, we denote $f_{ij}$ for $f (q_{ij},K_{ij})$ and 
$\phi_{ij}$ for $\phi (q_{ij},K_{ij})$. 
Equation (\ref{eq:c2}) shows that the two-particle correlation function is 
independent of $\phi_{ij}$.

Similarly, the three-particle spectrum and correlation are given by
\begin{eqnarray}
W_3({\bf{p}}_1,{\bf{p}}_2,{\bf{p}}_3) &=& \langle
   a^\dagger_{\rm{out}}({\bf{p}}_1) a_{\rm{out}}({\bf{p}}_1) 
   a^\dagger_{\rm{out}}({\bf{p}}_2) a_{\rm{out}}({\bf{p}}_2) 
   a^\dagger_{\rm{out}}({\bf{p}}_3) a_{\rm{out}}({\bf{p}}_3)  
   \rangle,
\end{eqnarray}
and
\begin{eqnarray}
\label{eq:c3}
C_3({\bf{p}}_1,{\bf{p}}_2,{\bf{p}}_3) 
&=& \frac{W_3({\bf{p}}_1,{\bf{p}}_2,{\bf{p}}_3)}
       {W_1({\bf{p}}_1)W_1({\bf{p}}_2)W_1({\bf{p}}_3)} \nonumber \\
 &=&  1 
 + \frac{f_{12}^2}{f_{11}f_{22}}
 + \frac{f_{23}^2}{f_{22}f_{33}}
 + \frac{f_{31}^2}{f_{33}f_{11}}  \nonumber \\
 && + 2 F_3 \cos\Phi,
\end{eqnarray}
where 
\begin{eqnarray}
\label{eq:Phig}
\Phi &=& \phi_{12}+\phi_{23}+\phi_{31} \nonumber \\
 F_3 &=& \frac{f_{12}f_{23}f_{31}}{f_{11}f_{22}f_{33}}.
\end{eqnarray}
From the preceding discussion of $\phi_{ij}$, we have 
\begin{equation}
\Phi (p_1=p_2=p_3) = 0
\end{equation}
for a real source function.  
When $\langle x \rangle_{K_{12}} \approx \langle x \rangle_{K_{23}} \approx 
\langle x \rangle_{K{31}}$,  Eq. (6) yields 
\begin{equation}
\Phi \approx  -\frac{1}{6}[
 \left\langle (q_{12} \cdot {\tilde{x}}_{K_{12}} )^3 \right\rangle_{K_{12}}
+\left\langle (q_{23} \cdot {\tilde{x}}_{K_{23}} )^3 \right\rangle_{K_{23}}
+\left\langle (q_{31} \cdot {\tilde{x}}_{K_{31}} )^3 \right\rangle_{K_{31}}]
\ ;
\label{mexphi}
\end{equation}
$\Phi$ is thus expected to vary prominently when the source is 
asymmetric (in $x$). 

Equation (\ref{eq:c3}) shows that 
the three-particle correlation function depends on 
$\cos\Phi$, which is absent in the two-particle
correlation.  The rest of the three-particle correlation is expressed 
in terms of $f_{ij}$'s and can thus be determined from 
the two-particle correlations.

The three-particle correlation is a function of 
three momenta, ${\bf{p}}_1, {\bf{p}}_2$, and ${\bf{p}}_3$.
For convenience, we introduce the average total momentum, 
\begin{equation}
K = \frac{p_1+p_2+p_3}{3}. 
\end{equation}
Since ${q}_{ij}$'s satisfy the identity,
\begin{equation}
{q}_{12}+{q}_{23}+{q}_{31} = 0,
\end{equation}
we have three independent momentum variables, which we will take to be 
${K}$, ${q}_{12}$, and ${q}_{23}$.

Among various choices of three variables, another convenient choice is
a set of ${K}_{ij}$'s.  For completeness, we show the relations 
between the first set and ${K}_{ij}$'s: 
\begin{equation}
{K}_{12}= {K}
               +\frac{1}{6}({q}_{23}-{q}_{31}), \ \ \ \ 
{K}_{23}= {K}
               +\frac{1}{6}({q}_{31}-{q}_{12}), \ \ \ \ 
{K}_{31}= {K}
               +\frac{1}{6}({q}_{12}-{q}_{23}),
\end{equation}
satisfying 
\begin{equation}
{K} = \frac{{K}_{12}+{K}_{23}+{K}_{31}}{3}.
\end{equation}

\section{Source function}

We apply the Bjorken hydrodynamical model to describe the evolution of 
the hot region created in heavy-ion collisions.
Based on the Cooper-Frye spectrum, one writes the source function as\cite{Heinz} 
\begin{eqnarray}
S(x,K) d^4x &=& F(\tau,\eta)d\tau \exp\left(-\frac{{\bf{x}}_T^2}{2R_T^2}\right)
   \times K \cdot d\sigma \exp\left(-\frac{K \cdot U}{T_f}\right) \nonumber \\
   &=& F(\tau,\eta)d\tau \exp\left(-\frac{{\bf{x}}_T^2}{2R_T^2}\right)
    \times m_T\cosh(Y-\eta)e^{-\frac{m_T}{T_f}\cosh(Y-\eta)} 
    \tau d\eta d^2{\bf{x}}_T,
\end{eqnarray}
where $\tau$, $\eta$, and ${\bf{x}}_T$ denote the coordinate variables defined 
as $\tau = \sqrt{t^2-z^2}, \eta = \frac{1}{2}\log\left(\frac{t+z}{t-z}\right)$,
and ${\bf{x}}_T = (x,y)$, respectively;  
$m_T$ and $Y$ are the momentum variables defined as $m_T = \sqrt{K_0^2-K_z^2}$ 
and $Y = \frac{1}{2}\log\left(\frac{K_0+K_z}{K_0-K_z}\right)$, respectively.
Note that we specify the momentum variables for a momentum pair, 
$p_i$ and $p_j$,  
by the subscript $ij$, such as $m_{Tij}$, and those for three 
momenta with no subscript such as $m_T$. 
We set the $z$-axis to be the beam direction, and the $x$-axis parallel
to ${\bf{K}}_T$, or perpendicular to the beam axis.
$U^\mu$ is a four-velocity of flow, and is 
$U^\mu = (\frac{t}{\tau},0,0,\frac{z}{\tau})$ 
in the Bjorken hydrodynamical model. 
The profile function 
$F(\tau,\eta)$ determines the source shape in the $\tau$-$\eta$ space.
The shape along the transverse direction is taken to be of a Gaussian form.
$T_f$ is the freeze-out temperature, and $d\sigma_\mu$ is the measure
of freeze-out hypersurface. We assume that freeze-out occurs
on the hypersurface where $\tau$ is constant.

In this work, we examine the following five forms of the profile function.
\\
Simple (box-type) profile:
\begin{equation}
F(\tau,\eta) = \left\{
\begin{array}{lc}
\delta(\tau-\tau_0) &
( \left|\eta \right|< \Delta\eta )\\
0 & (\mbox{otherwise})
\end{array}
\right.
\end{equation}
Gaussian profile:
\begin{equation}
F(\tau,\eta) = \delta(\tau-\tau_0) 
          \exp\left( -\frac{\eta^2}{2\Delta\eta^2} \right)
\end{equation}
Heinz profile:
\begin{equation}
F(\tau,\eta) = \frac{1}{(2\pi)\Delta\tau}
    \exp\left(-\frac{(\tau-\tau_0)^2}{2\Delta\tau^2}
      -\frac{\eta^2}{2\Delta\eta^2} \right)
\end{equation}
Exponential profile:
\begin{equation}
F(\tau,\eta) =  \frac{1}{\Delta\tau}\theta(\tau-\tau_0)
   \exp\left(-\frac{\tau-\tau_0}{\Delta\tau}
     -\frac{\eta^2}{2\Delta\eta^2} \right)
\end{equation}
Theta profile:
\begin{equation}
F(\tau,\eta) = \frac{1}{(\pi)\Delta\tau}\theta(\tau-\tau_0)
   \exp\left(-\frac{(\tau-\tau_0)^2}{2\Delta\tau^2}
   -\frac{\eta^2}{2\Delta\eta^2} \right)
\end{equation}

\section{Optimal variables}

The three-particle correlation function depends on three momenta, 
which have nine components.  As noted in Sec. II, we choose the three momenta 
to be $K$, $q_{12}$, and $q_{23}$.  In this work, we focus on the dependence of 
the correlation functions on relative momenta of the emitted bosons by fixing 
the value of $K$.  This leaves the two relative momenta to be 
the remaining variables.

\begin{figure}[p]
\begin{center}
\epsfig{file=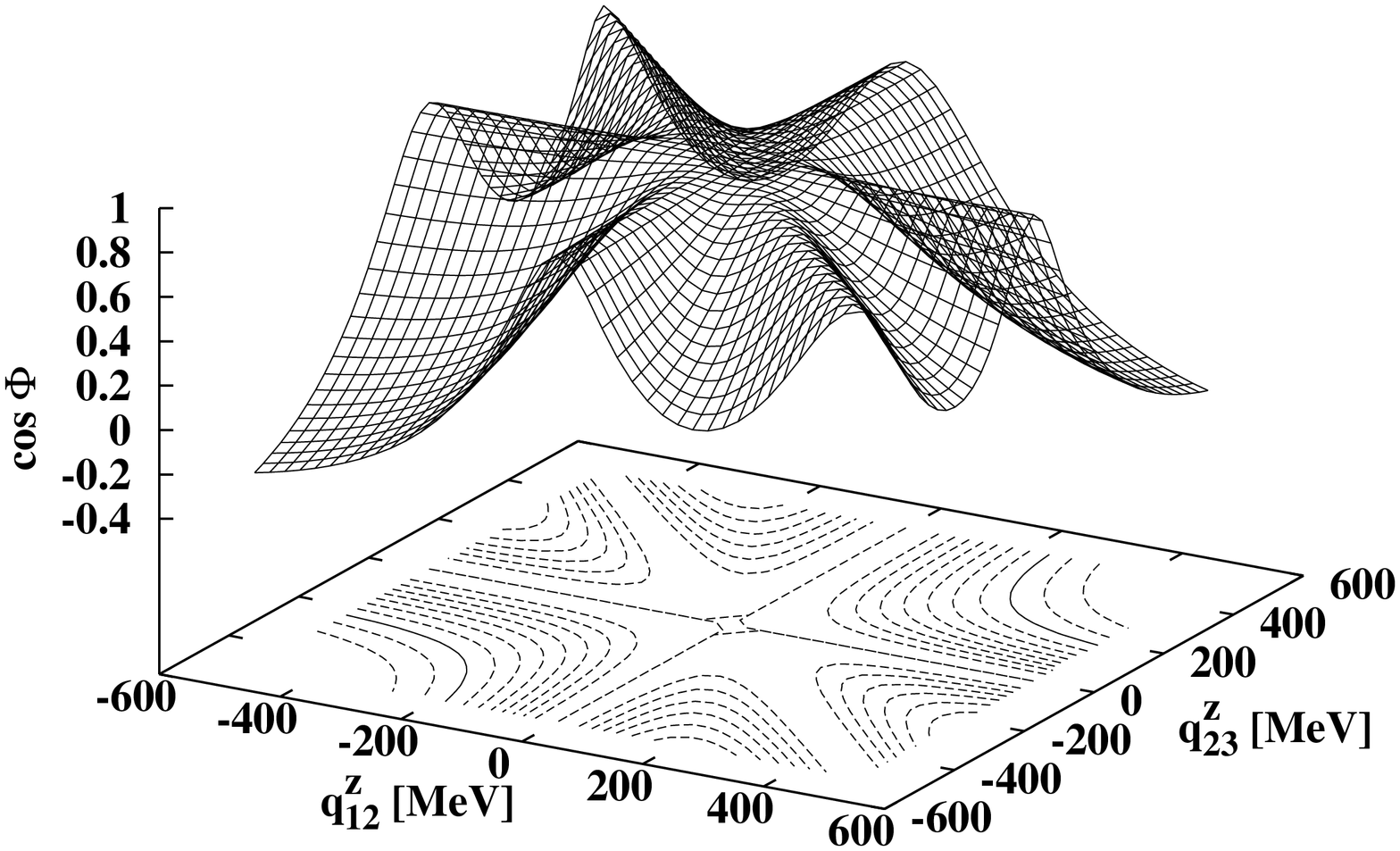,scale=0.5,angle=-90}
\end{center}
\caption{$\cos\Phi$ as a function of $q_{12}^z$ and $q_{23}^z$ 
 for the Heinz profile function.  The contour curves 
 on the $q_{12}^z - q_{23}^z$ plane are shown for the value of 
 $\cos\Phi$ from 0 to 1 by the step of 0.1. See the text for 
 the parameter values used.}
\label{fig-3d}
\end{figure}

In order to identify new information in the correlation function, we should 
choose the variables that could provide the most rapid variation 
of $\cos\Phi$.  Figure \ref{fig-3d} illustrates the variation of $\cos\Phi$
as a function of $q_{12}^z$ and $q_{23}^z$ with all other components of 
the relative momenta set to be zero.
Here, we use the Heinz profile 
with the parameters of $\tau_0 = 6$ fm, $\Delta\tau = 1$ fm,
$\Delta\eta = 1.2$, and $m_T = T_f = 140$ MeV.   Figure \ref{fig-3d} shows 
that $\cos\Phi$ is unity along the $q_{12}^z$- and $q_{23}^z$-axes
and also along the line of  $q_{12}^z = - q_{23}^z$, or the $q_{31}^z$-axis 
(because of Eq. (16)).  Figure \ref{fig-3d} also shows that $\cos\Phi$
varies most prominently along the lines of $q_{12}^z = q_{23}^z$, 
$q_{23}^z = q_{31}^z$, and $q_{31}^z = q_{12}^z$.  Because of symmetry, 
however, it is sufficient to examine the variation around one of the lines, 
such as the line of $q_{12}^z = q_{23}^z$.  

The reason why the line of $q_{12}^z = q_{23}^z$ provides a prominent variation 
of $\cos\Phi$ is seen as follows.  When $q_{ij} \ll K_{ij}$ and 
$K_{12} \approx K_{23} \approx K_{31}$, we have 
\begin{equation}
\Phi \approx \phi(q_{12}) + \phi(q_{23}) + \phi(-q_{12}-q_{23}) , 
\end{equation}
where the (weak) dependence on $K_{ij}$ in $\phi$ is not explicitly shown. 
Setting $q^z_{12} = q_z \cos\theta$ and $q^z_{23} = q_z \sin\theta$, we find 
\begin{eqnarray}
\frac{\partial\Phi} {\partial\theta} \approx 
&-&q_z \sin\theta \phi^{\prime}(q^x_{12},q^y_{12},q_z\cos\theta)
 +q_z \cos\theta \phi^{\prime}(q^x_{23},q^y_{23},q_z\sin\theta) \nonumber \\
&-&q_z (\cos\theta - \sin\theta) 
\phi^{\prime}(-q^x_{12}-q^x_{23},-q^y_{12}-q^y_{23},
              -q_z(\cos\theta+\sin\theta)),
\end{eqnarray}
where $\phi^{\prime}({\bf q}) = \partial\phi / \partial q_z$.  
When $q^x_{12}=q^x_{23}$ and $q^y_{12}=q^y_{23}$, 
$\partial\Phi / \partial\theta$ vanishes at $\theta = \pi/4$, 
or $q^z_{12}=q^z_{23}$,  yielding the minimum.  It follows then  
that the optimum choice is $q_{12} = q_{23}$.  

With this 
constraint, the variables are now reduced to a single relative momentum.

The source functions under consideration (as listed in Sec. III) are almost 
symmetric about the beam and transverse axes, but they can be asymmetric 
about the time-axis.  
Which component of $q_{12} ( = q_{23})$ should we choose 
so as to describe 
most effectively the asymmetric property of the source functions?    
$q^0_{12}$ is an obvious choice, but we find that $q^z_{12}$ or $q^x_{12}$ 
is the most convenient variable.  
From Eq. (10), we have $K_{12} \cdot q_{12} = 0$, which gives 
$q_{12} \cdot x = q^z_{12} (t K^z_{12}/K^0_{12}  - z) 
+ q^x_{12} (t K^T_{12}/K^0_{12} -x) - q^y_{12} y$.  
Thus $\cos\Phi$ as a function of $q^z_{12}$ or $q^x_{12}$ reflects time-axis 
asymmetry.  Furthermore, for finite $Y_{12}$, our source functions are not 
simply a function of $z^2$ without symmetry along the longitudinal direction.  
Because of this, $q^x_{12}$ may be the variable more suitable 
for identifying the time-axis 
asymmetry of the source function.  In the following, we will examine 
both $q^z_{12}$ and $q^x_{12}$.

$q^x_{12}$ is parallel to ${\bf{K}}^T_{12}$, while $q^y_{12}$ is perpendicular 
to it.  Following the practice in the literature on two-pion interferometry, 
we will also denote $q^x_{12}$ and $q^y_{12}$ as $q_{\rm{out}}$ and 
$q_{\rm side}$, respectively.  
Note that when the dependence of $\cos\Phi$ on $q^z_{12}$ or $q^x_{12}$ 
is examined, the other components of $q_{12}$ are fixed in our calculation.

\section{Numerical results}

\begin{figure}[p]
\begin{center}
\epsfig{file=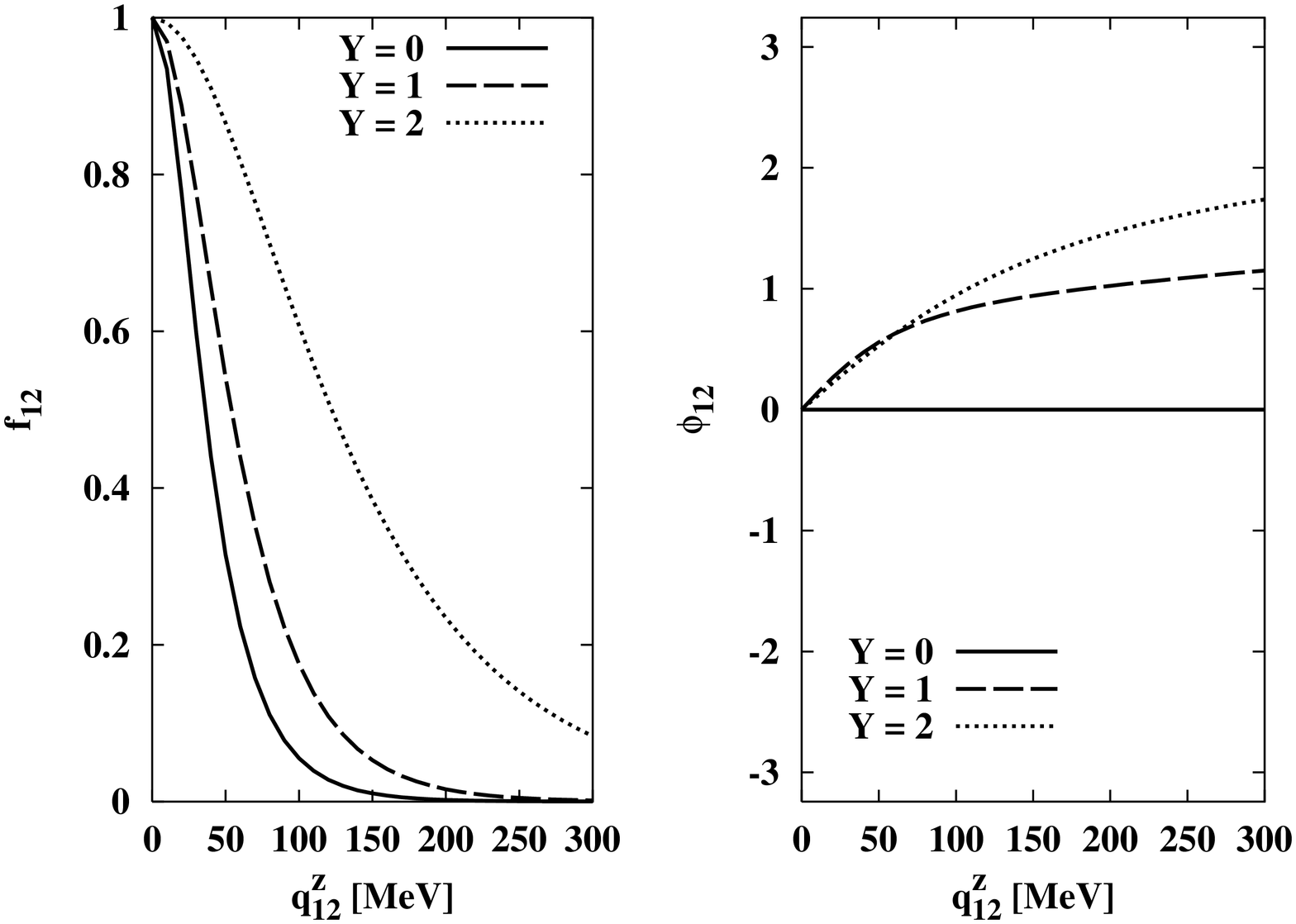,scale=0.5,angle=270}
\end{center}
\caption{$f_{12}$ and $\phi_{12}$ as functions of $q_{12}^z$ 
for $Y_{12} = 0, 1$, and 2 in the Heinz profile function.  See the text for 
 the parameter values used.}
\label{fig-h2zy}
\end{figure}

Figures \ref{fig-h2zy}, \ref{fig-s2zy}, \ref{fig-g2zy}, \ref{fig-t2zy},
and \ref{fig-e2zy} illustrate the dependence of $f_{12}$ and $\phi_{12}$ 
on $q_{12}^z$ in the Heinz, Simple, Gaussian, Theta, and Exponential profile 
functions, respectively.  Each figure is shown for $Y_{12}=0, 1$, and 2.  
In the figures, we set $q_{12}^{\rm out}=q_{12}^{\rm side}=0$, 
$m_{T12}=$ 140 MeV, $T_{f} =$ 140 MeV, $\Delta\eta = 1.2$, and $\tau_0=6$ fm 
for all profile functions, 
and also $\Delta\tau=1$ fm for the Heinz, Theta, and Exponential 
profile functions.  These parameter values reasonably satisfy the recent 
CERN-SPS experiment on Pb + Pb at 158 GeV/A\cite{NA49}.
Note that $|f_{12}|^2 + 1$ describes the two-particle correlation 
since $f_{12}$ is normalized to be unity at $q_{12}^z = 0$, and that 
$\phi$ is not observed in the experiment.

\begin{figure}[p]
\begin{center}
\epsfig{file=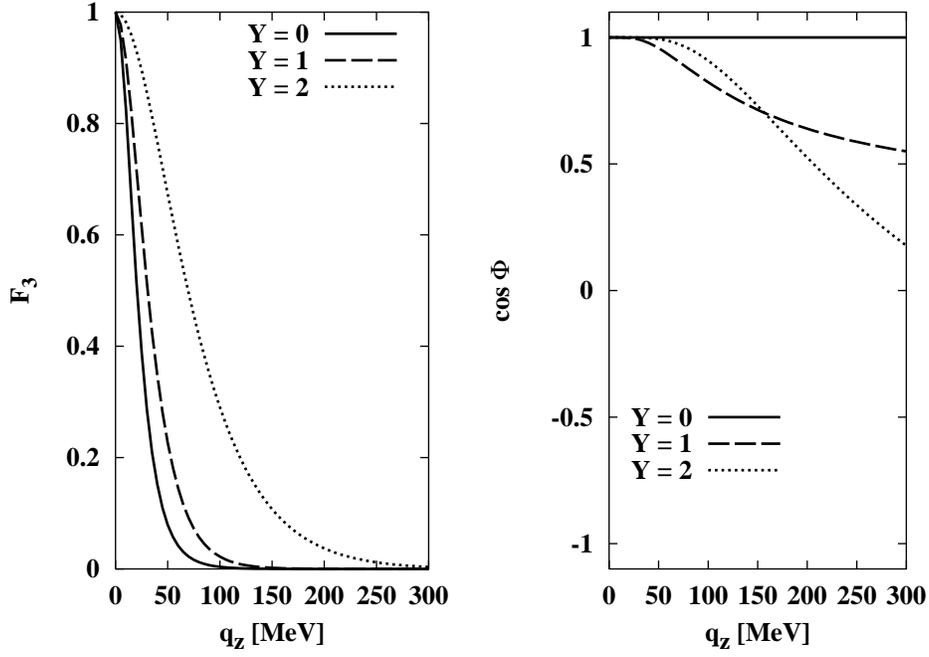,scale=0.5,angle=270}
\end{center}
\caption{$\cos\Phi$ and $F_3$ as functions of $q_z$ 
 with $Y = 0, 1$, and 2 for the Heinz profile function.
 See the text for the parameter values used.}
\label{fig-h3zy}
\end{figure}

\begin{figure}[p]
\begin{center}
\epsfig{file=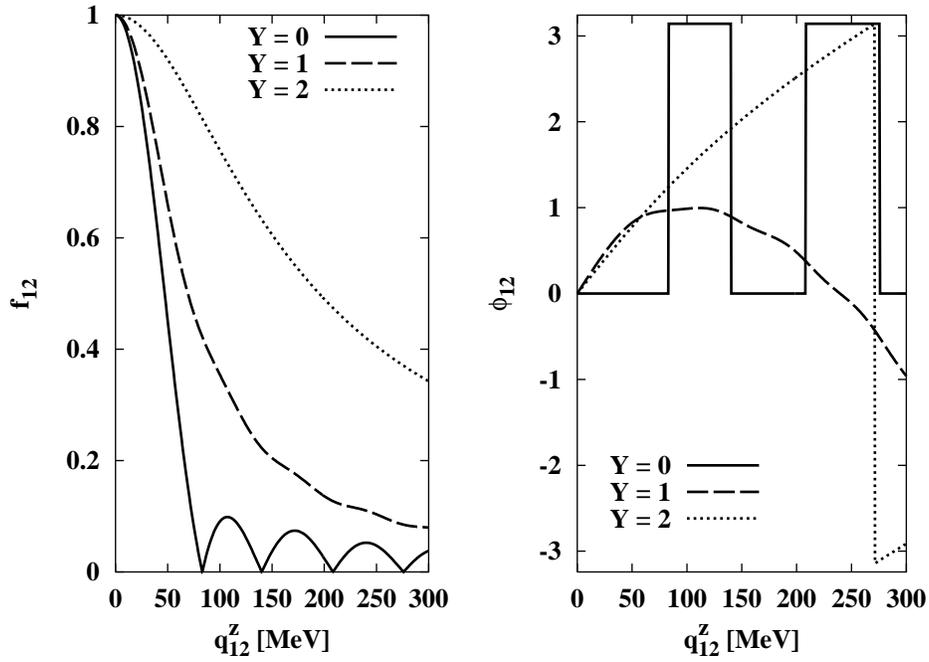,scale=0.5,angle=270}
\end{center}
\caption{$f_{12}$ and $\phi_{12}$ as functions of $q_{12}^z$ 
 with $Y_{12} = 0, 1$, and 2 for the Simple profile function.
 See the text for the parameter values used.}
\label{fig-s2zy}
\end{figure}

\begin{figure}[p]
\begin{center}
\epsfig{file=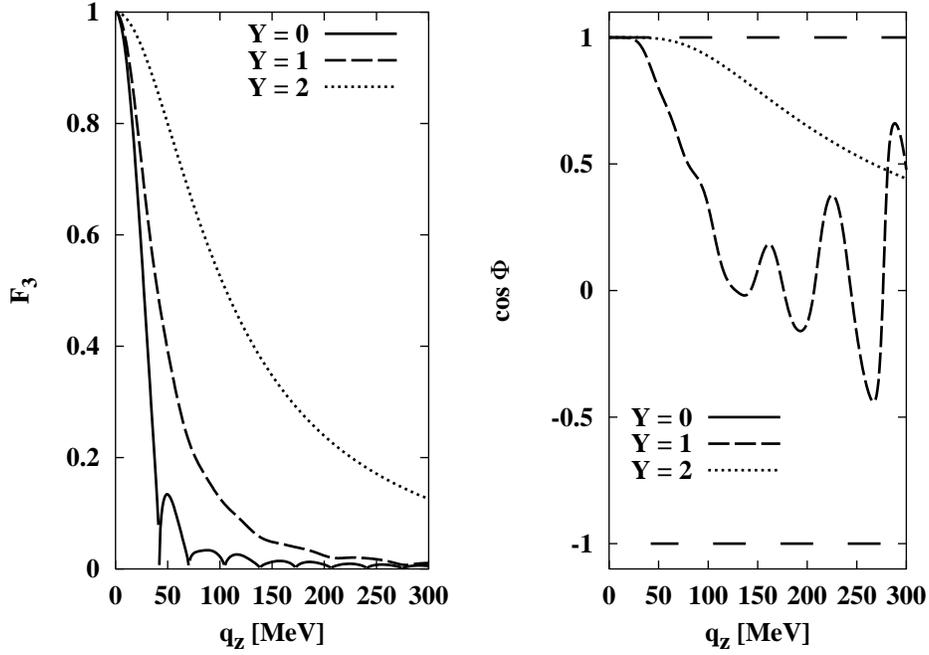,scale=0.5,angle=270}
\end{center}
\caption{$\cos\Phi$ and $F_3$ as functions of $q_z$ 
 with $Y = 0, 1$, and 2 in the Simple profile function.
 See the text for the parameter values used.}
\label{fig-s3zy}
\end{figure}

\begin{figure}[p]
\begin{center}
\epsfig{file=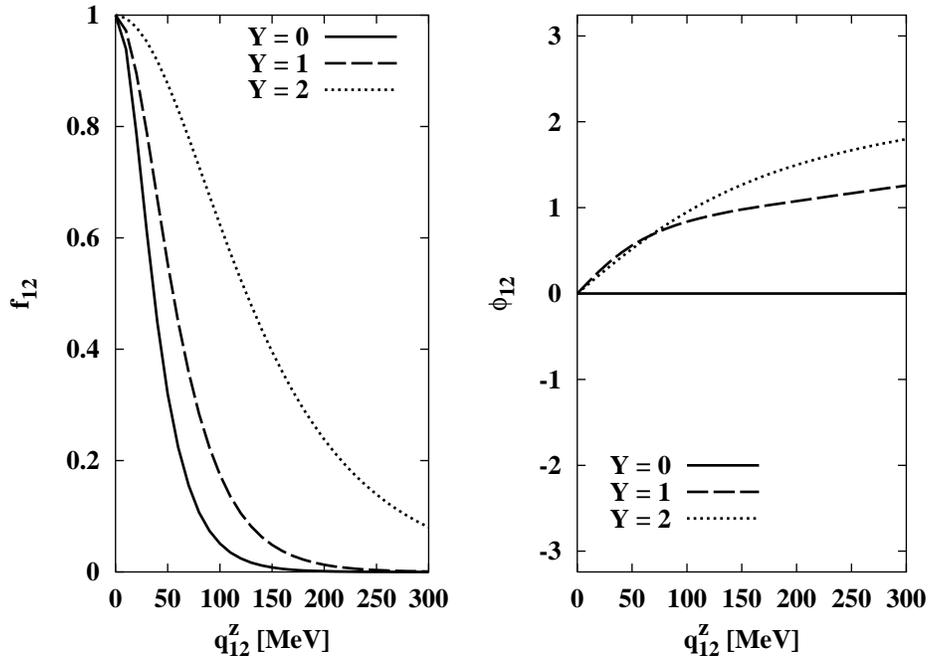,scale=0.5,angle=270}
\end{center}
\caption{$f_{12}$ and $\phi_{12}$ as functions of $q_{12}^z$ 
 with $Y_{12} = 0, 1$, and 2 in the Gaussian profile function.
 See the text for the parameter values used.}
\label{fig-g2zy}
\end{figure}

\begin{figure}[p]
\begin{center}
\epsfig{file=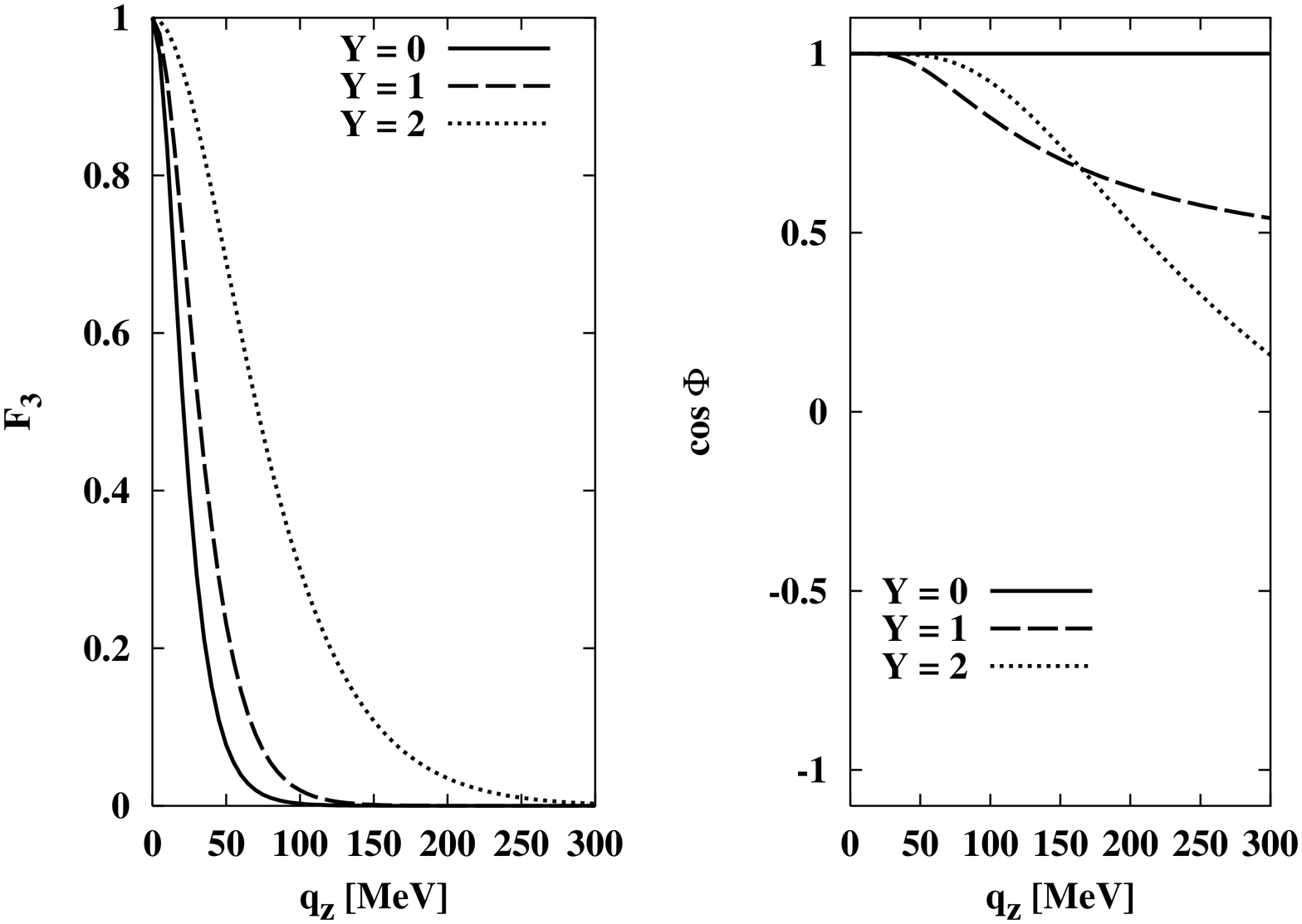,scale=0.5,angle=270}
\end{center}
\caption{$\cos\Phi$ and $F_3$ as functions of $q_z$ 
 for $Y = 0, 1$, and 2 in the Gaussian profile function.
 See the text for the parameter values used.}
\label{fig-g3zy}
\end{figure}

\begin{figure}[p]
\begin{center}
\epsfig{file=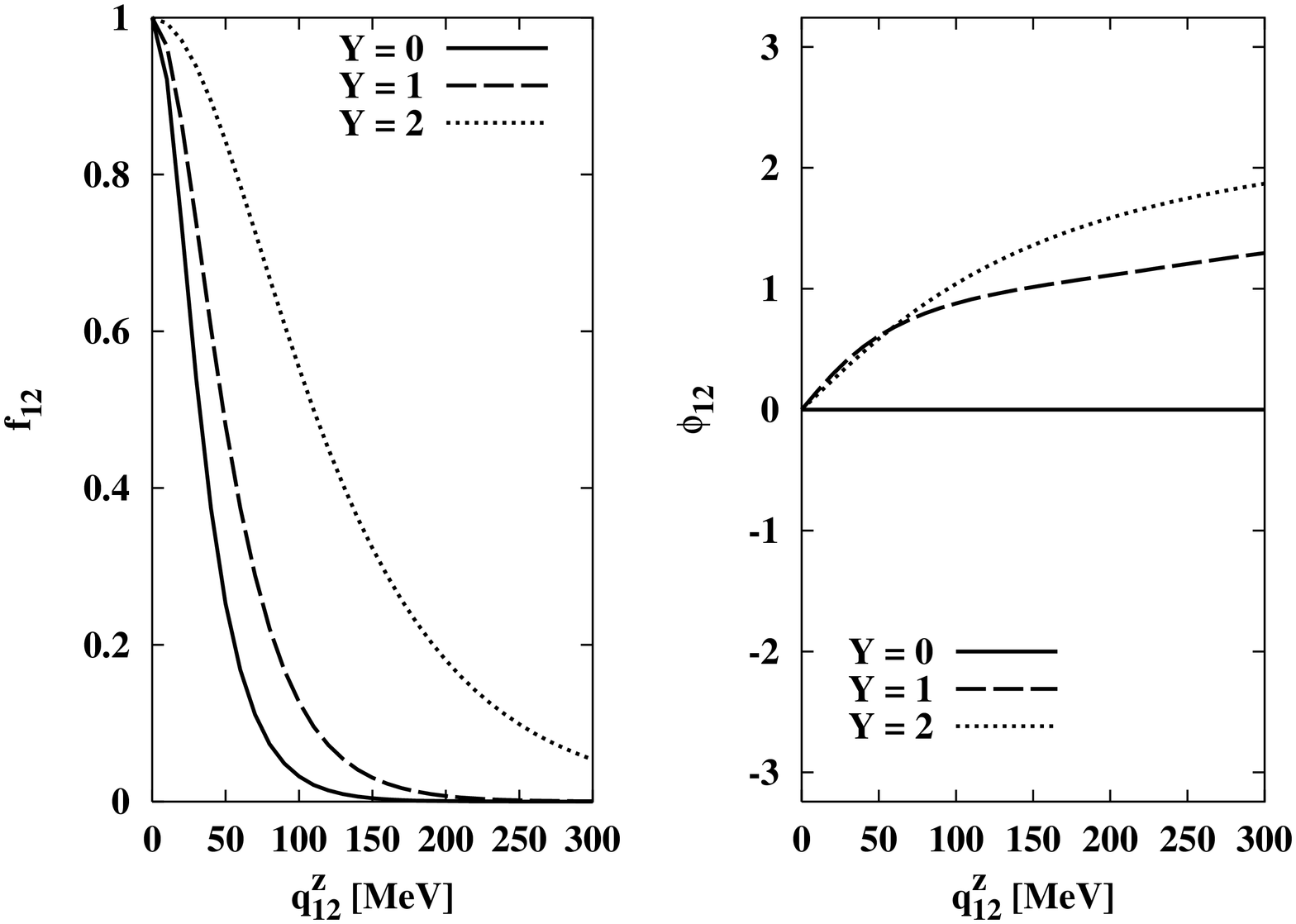,scale=0.5,angle=270}
\end{center}
\caption{$f_{12}$ and $\phi_{12}$ as functions of $q_{12}^z$
 for $Y_{12} = 0, 1$, and 2 in the Theta profile function.
 See the text for the parameter values used.}
\label{fig-t2zy}
\end{figure}

\begin{figure}[p]
\begin{center}
\epsfig{file=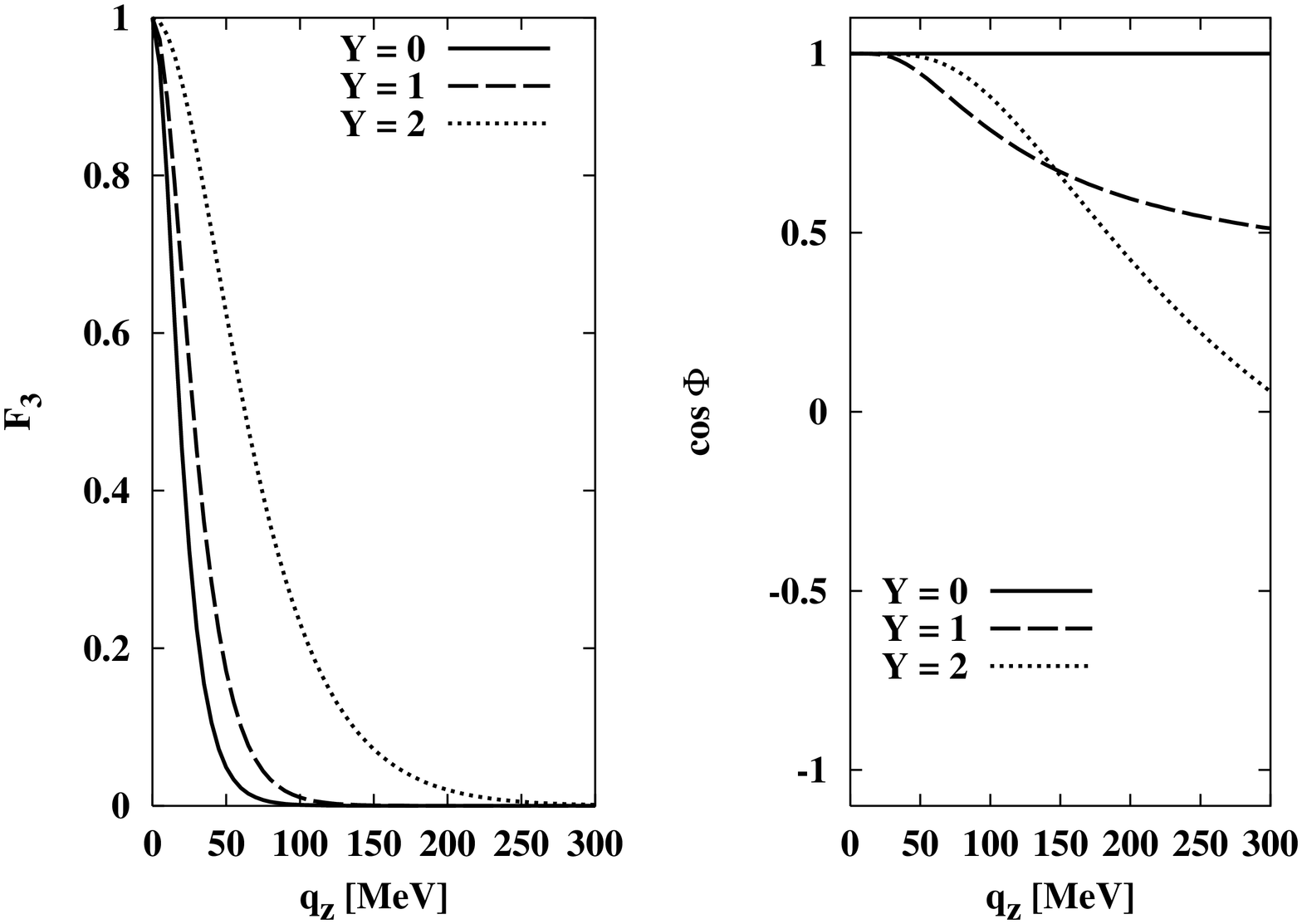,scale=0.5,angle=270}
\end{center}
\caption{$\cos\Phi$ and $F_3$ as functions of $q_z$ 
 for $Y = 0, 1$, and 2 in the Theta profile function. 
 See the text for the parameter values used.}
\label{fig-t3zy}
\end{figure}

\begin{figure}[p]
\begin{center}
\epsfig{file=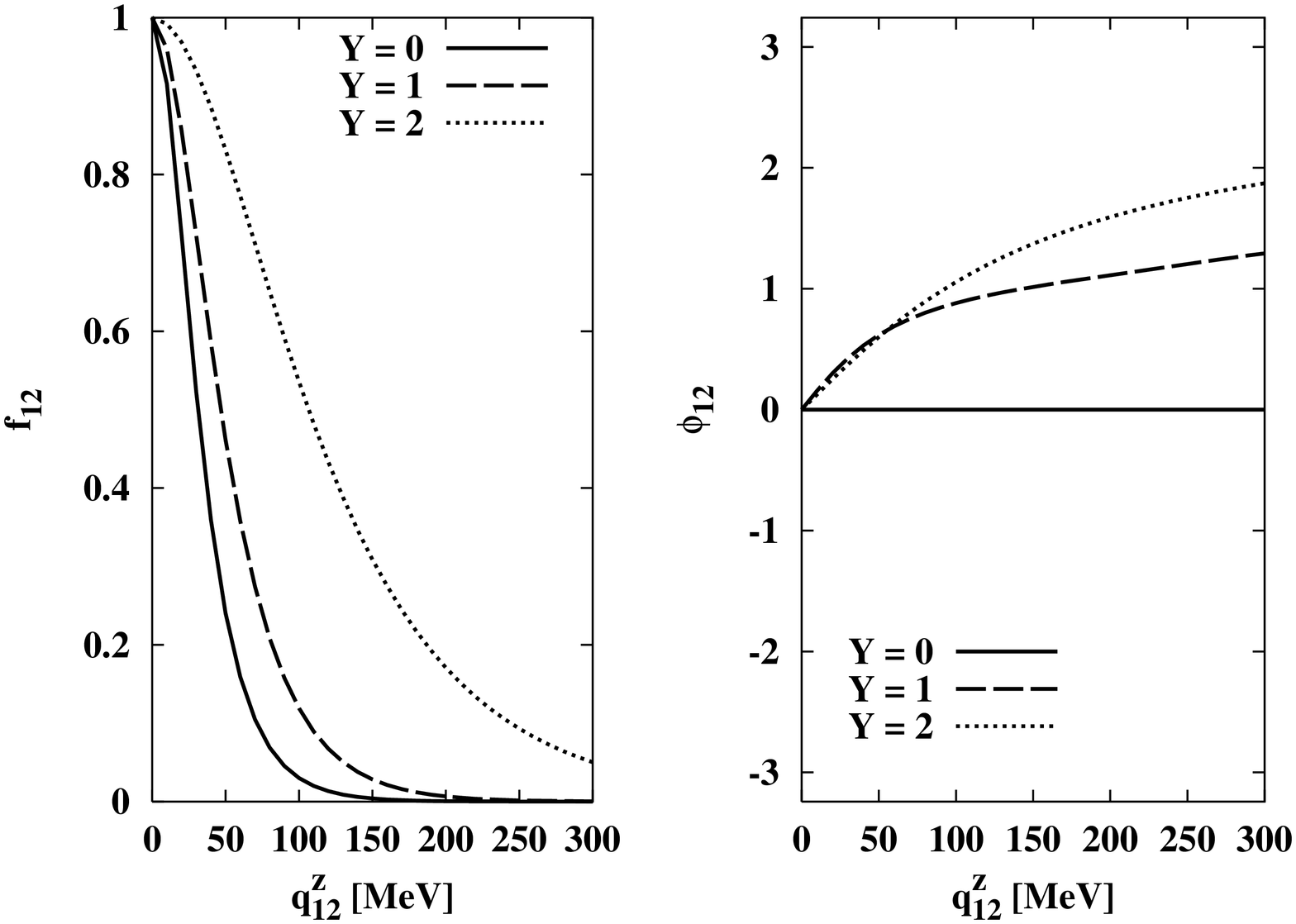,scale=0.5,angle=270}
\end{center}
\caption{$f_{12}$ and $\phi_{12}$ as functions of $q_{12}^z$
 for $Y_{12} = 0, 1$, and 2 in the Exponential profile function.
 See the text for the parameter values used.}
\label{fig-e2zy}
\end{figure}

\begin{figure}[p]
\begin{center}
\epsfig{file=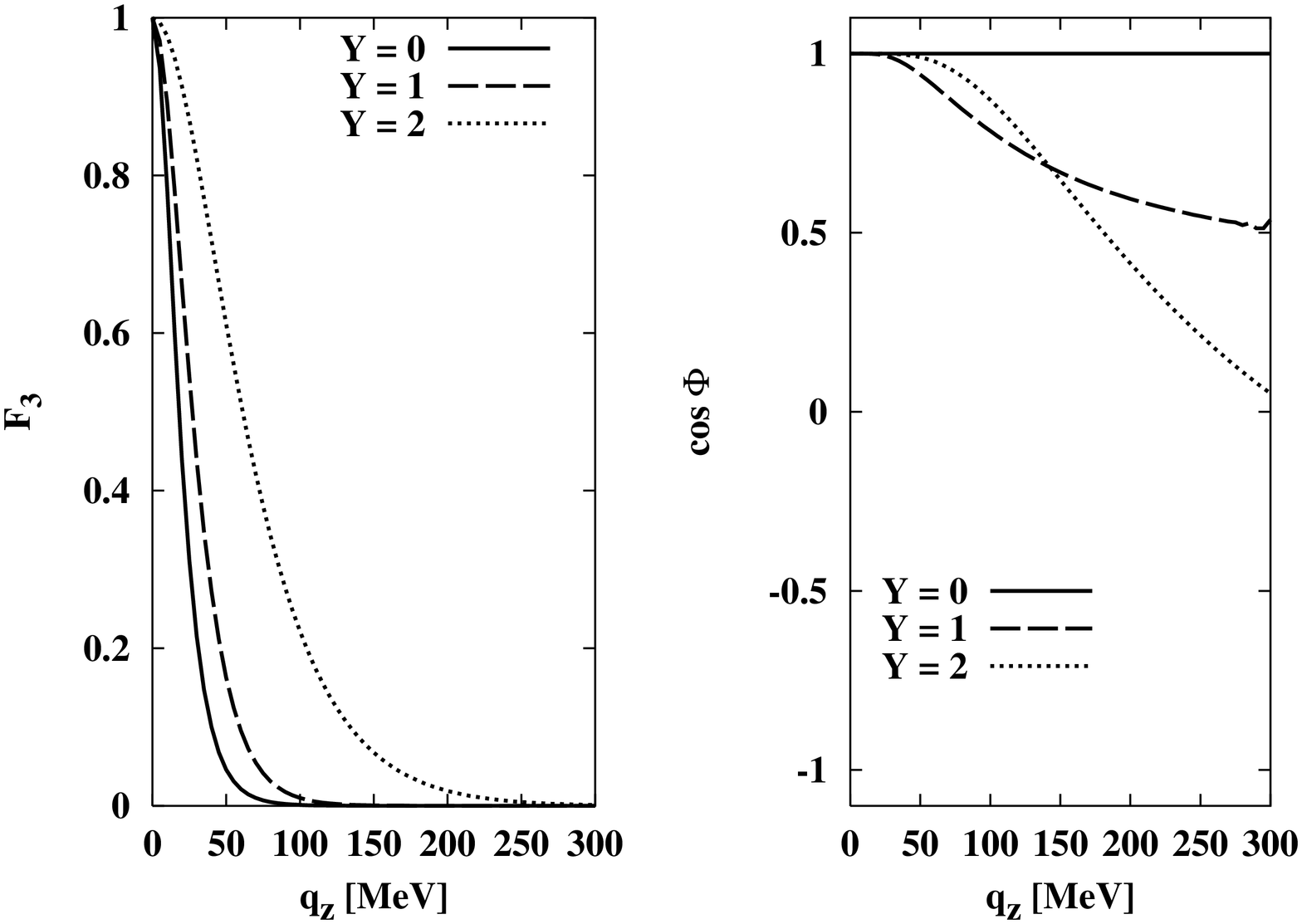,scale=0.5,angle=270}
\end{center}
\caption{$\cos\Phi$ and $F_3$ as functions of $q_z$ 
 for $Y = 0, 1$, and 2 in the Exponential profile function.
 See the text for the parameter values used.}
\label{fig-e3zy}
\end{figure}

Figures \ref{fig-h3zy}, \ref{fig-s3zy}, \ref{fig-g3zy}, \ref{fig-t3zy}, 
and \ref{fig-e3zy}
show $\cos\Phi$ and (half of) its coefficient in the three-particle 
correlation function, $F_3$, 
as functions of $q_z$ ( $\equiv q_{12}^z = q_{23}^z$ ).
Each figure is shown for $Y=0,1,$ and 2 and $m_T =$ 140 MeV 
with the parameter values of $T_f$, $\Delta\eta$, $\tau_0$, and $\Delta \tau$  
as in Figs.~\ref{fig-h2zy}, \ref{fig-s2zy}, \ref{fig-g2zy}, \ref{fig-t2zy}, 
and \ref{fig-e2zy}.

In Figs. \ref{fig-h3zy}, \ref{fig-s3zy}, \ref{fig-g3zy}, \ref{fig-t3zy}, 
and \ref{fig-e3zy}, we see that for all profile functions, $\cos\Phi = 1$  
at $Y = 0$, and $\cos\Phi$ becomes more prominent as $Y$ increases.  
$\cos\Phi$ tends to be smaller for the asymmetric profile functions 
(such as the Simple and Exponential profile functions) than 
for the symmetric ones, though the difference between them is not substantial. 
$\Phi$ tends to deviate from zero more slowly than  
the phases of $\rho_{ij}$'s, $\phi_{ij}$'s.
For example, we see in Figs. 4 and 5 that when $\phi_{12}$ reaches $\pi/2$ 
around $q^z_{12} =$ 150 MeV, we still have $\cos\Phi > 0.5$. 
The reason for this is general because $\Phi$ is defined as 
$\Phi = \phi_{12}+\phi_{23}+\phi_{31}$ and is constrained by the identity, 
$q_{12}+q_{23}+q_{31} = 0$.  In fact, the small-momentum 
expansion of Eq. (\ref{exphi}) applied to $\Phi$ yields that the term linear 
in $q$ vanishes, as discussed in Sec. II.  Furthermore, when $\Phi$ starts 
to deviate from zero, its coefficient in the three-particle correlation 
function, $2F_3$, tends to become small.  In fact, we see in the figures 
that $2F_3$ gets halved for all profile functions 
before $\cos\Phi$ decreases to 0.9, and even nearly 
vanishes when $\cos\Phi$ decreases further.
We expect that in actual experiments it will be difficult to identify 
$\cos\Phi$ with a value much smaller than unity.   
This is the major finding of this work.

\begin{figure}[p]
\begin{center}
\epsfig{file=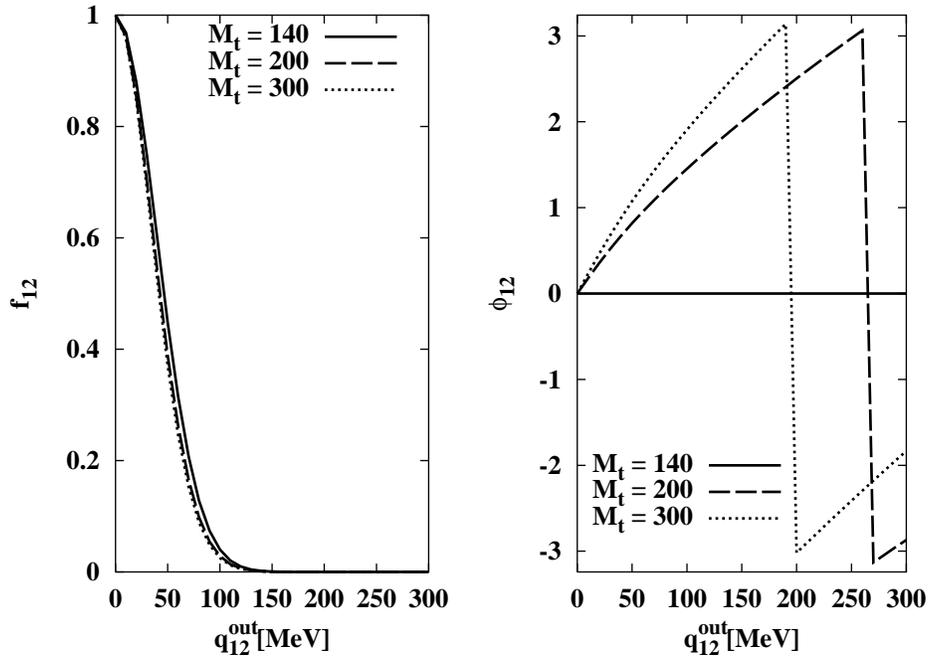,scale=0.5,angle=270}
\end{center}
\caption{$f_{12}$ and $\phi_{12}$ as functions of $q_{12}^{out}$ 
 for $m_{T12} = 140, 200$, and 300 MeV in the Heinz profile function.
 See the text for the parameter values used.}
\label{fig-h2tm}
\end{figure}

\begin{figure}[p]
\begin{center}
\epsfig{file=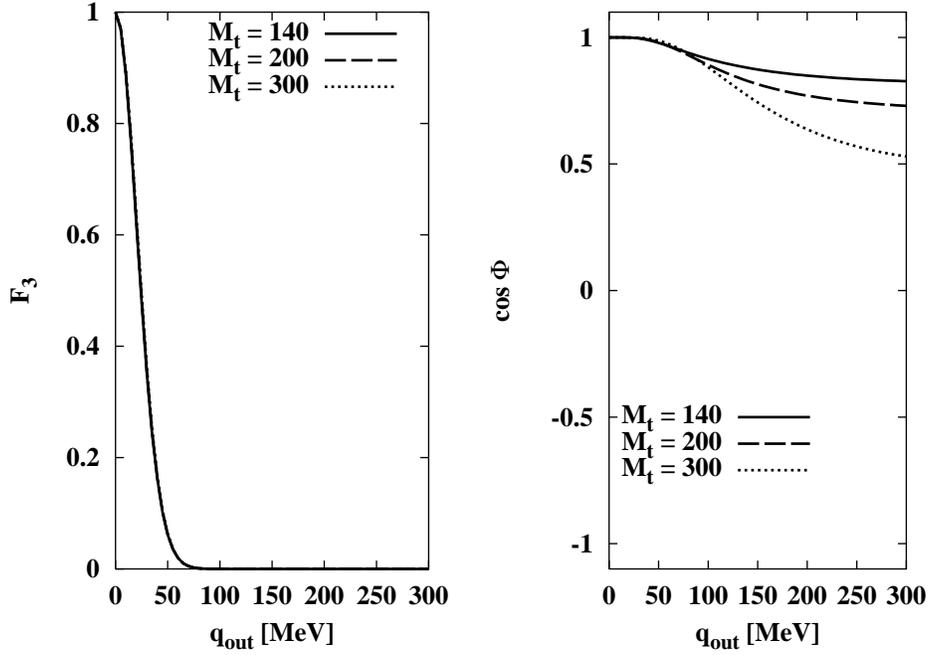,scale=0.5,angle=270}
\end{center}
\caption{$\cos\Phi$ and $F_3$ as functions of $q_{\rm{out}}$ 
 for $m_T = 140, 200$, and 300 MeV in the Heinz profile function.
 See the text for the parameter values used.}
\label{fig-h3tm}
\end{figure}

We find the same difficulty when we choose $q_{\rm out}$ as the independent 
variable.  Figure \ref{fig-h2tm}
shows $f_{12}$ and $\phi_{12}$ as functions of $q_{12}^{\rm out}$
for the Heinz profile for $m_{T12} =$ 140, 200, and 300 MeV,
with $\tau_0 = 6$ fm, $Y_{12} = 1.5$, $\Delta\eta = 1.2$, $\Delta\tau =1$ fm,
and $R_T = 5$ fm.  Figure \ref{fig-h3tm} also shows $\cos\Phi$ and $F_3$
as functions of $q_{\rm{out}} (= q_{12}^{out} = q_{23}^{out})$ 
for the three values of $m_T$ 
with $Y =$ 1.5 and the same values of $T_f$, $\Delta\eta$, $\Delta\tau$, 
$\tau_0$, and $R_T$. 
We see in Fig. \ref{fig-h3tm} that when $\cos\Phi$ starts to deviate 
from unity, its coefficient, $2 F_3$, becomes small quite rapidly, 
as in the previous cases of $q^z_{12}$ being the independent variable.  
Note that $F_3$ decreases more quickly as a function of $q_{\rm out}$ than 
of $q_z$, as seen in Figs. \ref{fig-h3tm} and \ref{fig-h3zy}. 
The quick reduction occurs because of the factor 
$\exp\left(-\frac{{\bf{x}}_T^2}{2R_T^2}\right)$, 
when the transverse momentum, $q_{\rm out}$, is used as the variable. 
$R_T=5$ fm is chosen to reproduce experimental results, but we find that 
even if a much smaller value of $R_T$ is used, $F_3$ becomes quite small 
when $\cos\Phi$ is off unity.

\begin{figure}[p]
\begin{center}
\epsfig{file=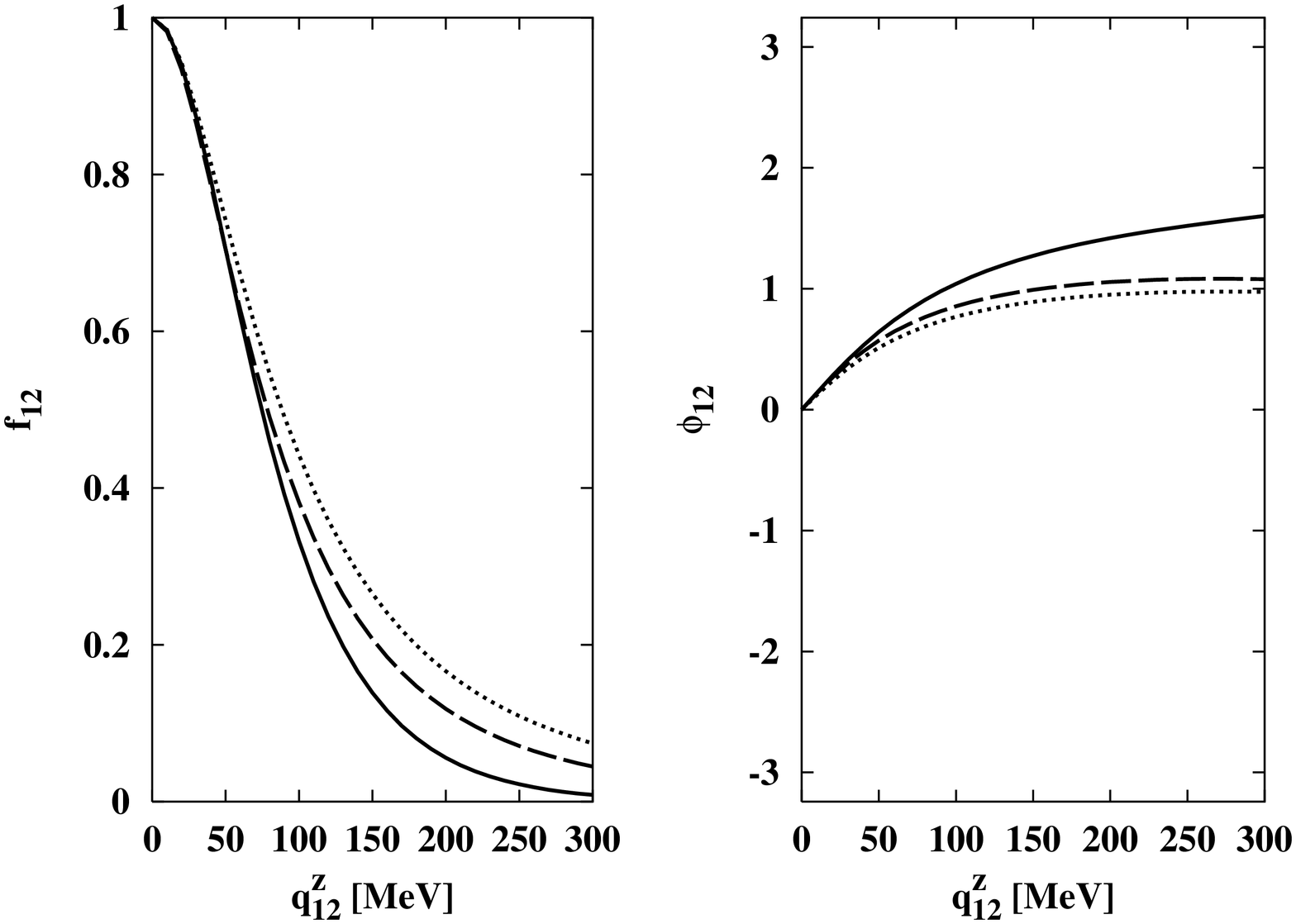,scale=0.5,angle=270}
\end{center}
\caption{$f_{12}$ and $\phi_{12}$ as functions of $q_{12}^z$ 
 for $(\tau_0, \Delta\tau) = (6.5, 0.65), (4.8,2.4)$, and (3.2,3.2) fm 
 for the Heinz profile function, shown by solid, dashed, and dotted 
 curves, respectively.
 See the text for the parameter values used.}
\label{fig-h2zdt}
\end{figure}

\begin{figure}[p]
\begin{center}
\epsfig{file=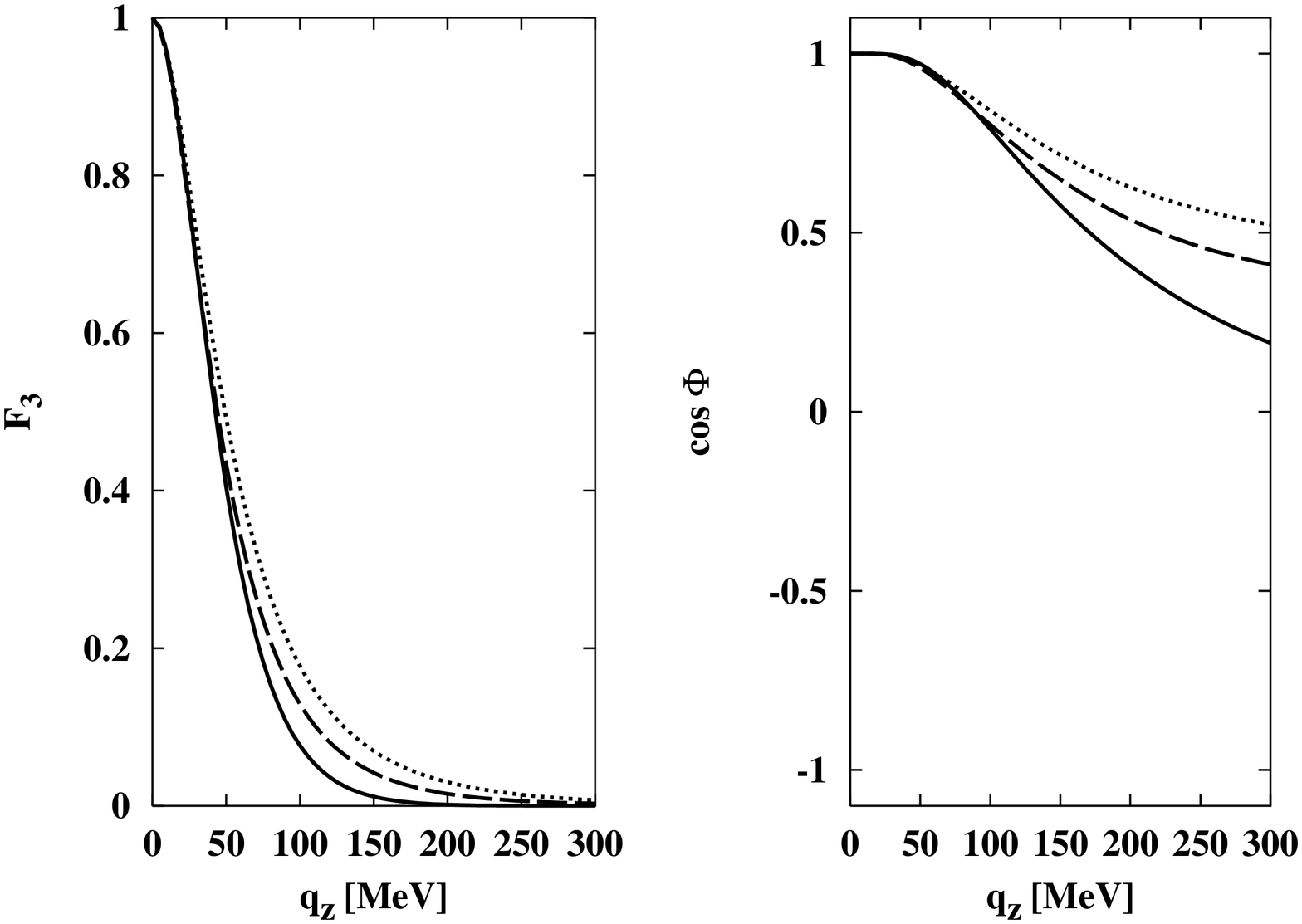,scale=0.5,angle=270}
\end{center}
\caption{$\cos\Phi$ and $F_3$ as functions of $q_z$ 
 for $(\tau_0, \Delta\tau) = (6.5, 0.65), (4.8,2.4)$, and (3.2,3.2) fm 
 for the Heinz profile function, shown by solid, dashed, and dotted 
curves, respectively.
 See the text for the parameter values used.}
\label{fig-h3zdt}
\end{figure}

The difficulty also remains at different values of $\tau_0$.  
Figures \ref{fig-h2zdt} and \ref{fig-h3zdt} show
$f_{12}$, $\phi_{12}$, $\cos\Phi$, and $F_3$ as functions of 
$q_{12}^z$ for $(\tau_0, \Delta\tau) =$ (6.5, 0.65), (4,8, 2.8), and 
(3.2, 3.2) fm, with $m_T = m_{T12} = 140$ MeV, $T_f = 140$ MeV, 
$Y = Y_{12} = 1.5$, and $\Delta\eta = 1.2$.
These parameter sets yield the recently measured value of the longitudinal 
size, $R_{\rm{long}}=$ 3.5 fm\cite{NA49}.
Figure \ref{fig-h3zdt} shows that $\cos\Phi$ depends on $\tau_0$ rather 
strongly.  But the variation of $\cos\Phi$ takes place where $F_3$ is small, 
and will be difficult to observe. 

\section{Discussions and summary}

We have investigated the three-particle correlation function for chaotic 
source, using various profile functions of the source in the Bjorken 
hydrodynamical model.  In all profile functions we have examined,  
the coefficient of $\cos\Phi$ in the three-particle correlation function, 
$2 F_3$, decreases quite rapidly, as $\cos\Phi$ decreases from unity.  
Extraction of $\cos\Phi$ is thus difficult when $\cos\Phi$ decreases from 
unity.  This result is in agreement with Heinz and Zhang\cite{HeiZha}, 
who previously made a similar investigation for small relative momenta.  
 
In order to clarify why $\cos\Phi$ is difficult to observe, we first identify 
the portion of the source function $S(x)$ that most strongly influences 
the phase, $\phi$.  $\phi$ appears in the Fourier transform of $S(x)$ as 
$\rho(q) = \rho_{R}(q) + i \rho_{I}(q) = f e^{i\phi}$.  As noted previously, 
$\phi$ is an odd function of $q$, and $\phi(q=0) = 0$ for a real source 
function.  The imaginary part of $\rho(q)$, $\rho_{I}(q)$, thus represents 
the behavior of $\phi$ relevant to the present discussion and is an odd function 
of $q$.  The portion of $S(x)$ that most strongly affects $\phi$ is thus its  
odd-function part, $S_{odd}(x)$, in $S(x) = S_{even}(x)+S_{odd}(x)$ with 
$S_{odd}(-x) = -S_{odd}(x)$ and $S_{even}(-x)=S_{even}(x)$.

\begin{figure}
\begin{center}
\epsfig{file=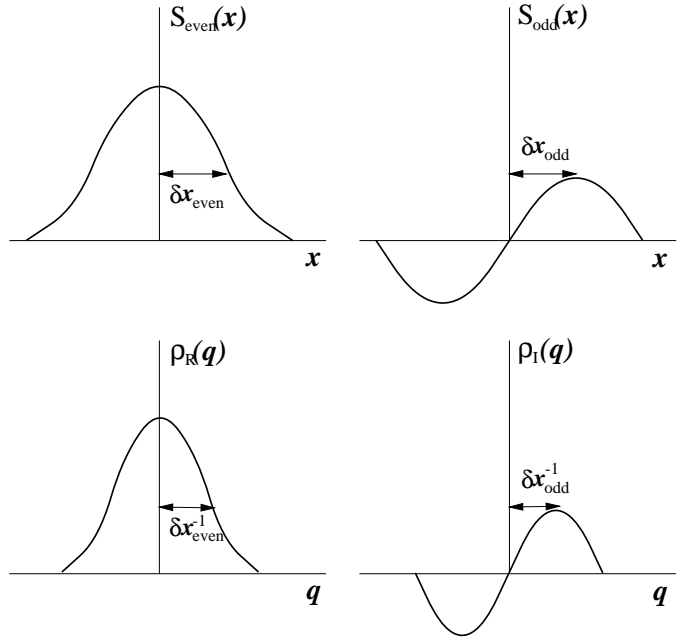,scale=0.5}
\end{center}
\caption{Distances of typical scale over which $S_{even}$ 
 and $S_{odd}$ vary.}
\label{fig-ex}
\end{figure}

Consider that $S_{even}(x)$ and $S_{odd}(x)$ vary over distances 
of typical scale, $\delta x_{even}$ and $\delta x_{odd}$, respectively. 
Here, $\delta x_{even}$ and $\delta x_{odd}$
correspond to the spread widths of $S_{even}(x)$ 
and the distance to maximum of $S_{odd}(x)$, respectively,  
as Fig. \ref{fig-ex} illustrates.
$\rho_{R}(q)$ and $\rho_{I}(q)$ also vary over the 
scales of $\delta x_{even}^{-1}$ and $\delta x_{odd}^{-1}$,
respectively.  As $\phi$ increases, $\rho_{I}(q)$ increases.  
The region where $\rho_{I}(q)$ is appreciable is $q > \delta x_{odd}^{-1}$.
The amplitude $f$ becomes small in the region where $q$ is larger than
$\delta x_{even}^{-1}$. 

Thus, if $\delta x_{odd}^{-1} < \delta x_{even}^{-1}$, the variation 
of $\phi$ is not difficult to observe.  This is, however, untenable 
as long as $S(x)$ is positive everywhere.  Consider 
$S_{odd}(x) < 0$ in some region:  we then have $S_{even} (x) > |S_{odd}(x)|$ 
in that region, and furthermore, $\delta x_{even} > \delta x_{odd}$.
The last inequality contradicts the desired inequality.

In this discussion, we have assumed the source function, $S(x,K)$, 
to be positive everywhere, as done in common practice.  
How realistic is this assumption?    
If $S(x,K)$ is a statistical phase-space distribution,
$S(x,K)$ should not be negative everywhere.  Generally speaking, however,   
$S(x,K)$ can be locally negative, provided its integrals over $x$ or $K$ be 
positive. This property is similar to, for example, that of the Wigner function.
Furthermore, $S(x,K)$ that is locally negative suggests that it would be 
associated with some dynamics involving quantum correlation.  This aspect of 
the interferometry is currently under investigation.

Measurement of the three-pion correlations from the CERN NA44 experiment 
has been recently reported\cite{exp} for the total relative momentum of up to 
about 300 MeV, and the extraction of $\cos\Phi$ up to about 60 MeV.  
This total relative momentum is too small 
to observe the variation of $\cos\Phi$ that is examined in this work. 
Taken as a constant, $\cos\Phi$ is found to be about 0.2\cite{exp} and is 
much smaller than the value of unity for a chaotic source, 
as discussed in this work.  The reason for the small value is unknown, but it 
may be because of partial coherence of the source, or even because of 
a more exotic reason\cite{dcc}.

In summary, we have investigated three-particle interferometry from 
chaotic sources using the Bjorken hydrodynamical model.
The optimal variables are identified suitable for extracting new 
information through the phase of the source function's Fourier transform.
The three-particle correlations are calculated over a wide range 
of the kinematic variables.  A variation of the three-pion phase is found 
to be difficult to observe experimentally because 
its coefficient, as it appears in the three-particle correlation function, 
becomes small in that region.  This is the case if the source function 
is positive everywhere as conventionally assumed, and suggests the 
interesting possibility of the source function being locally negative. 

\acknowledgements
We acknowledge M.~C.~Chu for his valuable contribution at the initial 
stage of this work.  
This research is partially supported by the U.S.~National Science
Foundation under grants PHY88-17296 and PHY90-13248 at Caltech, and 
the U.S.~Department of Energy under grant DE-FG03-87ER40347 at CSUN.

\end{document}